\documentclass[aps,prb,twocolumn,showpacs]{revtex4-1} 

\usepackage{graphicx}
\DeclareGraphicsExtensions{.png,.jpg,.eps,.pdf}

\usepackage{xcolor}
\usepackage{hyperref}
\usepackage{amsmath}

\usepackage{amssymb}
\usepackage{bm}
\usepackage{multirow}

\usepackage{amsthm}
\usepackage{bm}
\usepackage{amsfonts}
\usepackage{soul}

\usepackage{soul}

\usepackage[switch]{lineno} 

\newcommand{\misha}[1] {{\color{teal}\textbf{#1}}\color{black}\normalsize}

\newcommand{\beq}{\begin{equation}}  
\newcommand{\eeq}{\end{equation}}  
\newcommand{\beqa}{\begin{eqnarray}}  
\newcommand{\eeqa}{\end{eqnarray}}  

\let\oldthebibliography=\thebibliography
    \let\oldendthebibliography=\endthebibliography
    \renewenvironment{thebibliography}[1]{%
        \oldthebibliography{#1}%
        \setcounter{enumiv}{ 30 }%
    }{\oldendthebibliography} 

\newcounter{defcounter}
\newenvironment{myequation}{%
\addtocounter{equation}{-1}
\refstepcounter{defcounter}

\begin{equation}}
{\end{equation}}

\newenvironment{myequationa}{%
\addtocounter{equation}{-1}
\refstepcounter{defcounter}

\begin{eqnarray}}
{\end{eqnarray}}

\newcommand{\beqSM}{\begin{myequation}}  
\newcommand{\eeqSM}{\end{myequation}}  
\newcommand{\beqaSM}{\begin{myequationa}}  
\newcommand{\eeqaSM}{\end{myequationa}}

\begin{document}


\title{Spin wave excitations in low dimensional systems with large magnetic anisotropy}

\author{F. Delgado}
\address{Instituto de estudios avanzados IUDEA, Departamento de F\'{i}sica, 
Universidad de La Laguna, C./Astrof\'{i}sico Francisco S\'anchez, s/n. 38203, La Laguna}

\author{M.M. Otrokov}
\affiliation{Centro de F\'{i}ica de Materiales CFM/MPC (CSIC-UPV/EHU),
Paseo Manuel de Lardiz\'{a}bal 5, 20018 Donostia-San Sebasti\'{a}n, Spain}
\affiliation{Donostia International Physics Center, Paseo Manuel de Lardiz\'{a}bal 4, 20018 Donostia-San Sebasti\'{a}n, Spain}
\affiliation{IKERBASQUE, Basque Foundation for Science, E-48011 Bilbao, Spain}

\author{A. Arnau}
\affiliation{Centro de F\'{i}sica de Materiales CFM/MPC (CSIC-UPV/EHU),
Paseo Manuel de Lardiz\'{a}bal 5, 20018 Donostia-San Sebasti\'{a}n, Spain}
\affiliation{Donostia International Physics Center, Paseo Manuel de Lardiz\'{a}bal 4, 20018 Donostia-San Sebasti\'{a}n, Spain}
\affiliation{Departamento de Pol\'{i}meros y Materiales Avanzados: F\'{i}sica,
Qu\'{i}mica y Tecnolog\'{i}a, Facultad de Qu\'{i}mica UPV/EHU,
Apartado 1072, 20080 Donostia-San Sebasti\'{a}n, Spain}

\begin{abstract}
The low energy excitation spectrum of a two-dimensional ferromagnetic material is dominated by single-magnon excitations that show a gapless parabolic dispersion relation with the spin wave vector. This occurs as long as magnetic anisotropy and anisotropic exchange are negligible compared to isotropic exchange. However, to maintain magnetic order at finite temperatures, it is necessary to have sizable anisotropy to open a gap in the spin wave excitation spectrum. We consider four real two-dimensional systems for which ferromagnetic order at finite temperature has been observed or predicted.
Density functional theory calculations of the total energy differences for different spin configurations permit us to extract the relevant parameters 
and connect them with a spin Hamiltonian. 
The corresponding values of the Curie temperature are estimated using a simple model and found to be mostly determined by the value of the isotropic exchange. The exchange and anisotropy parameters are used in a toy model of finite-size periodic chains to study the low-energy excitation spectrum, including single-magnon and two-magnon excitations. 
At low energies we find that single-magnon excitations appear in the 
spectrum together with two-magnon excitations. These excitations present a gap that grows particularly for large values of the magnetic anisotropy or anisotropic exchange, relative to the isotropic exchange. 
\end{abstract}

%

\maketitle
%

\section{Introduction}
The appearance of ferromagnetic order at finite temperatures in two-dimensional systems requires the existence of magnetic anisotropy, so that a gap appears in the magnon (spin wave) excitation spectrum. Indeed, the magnitude of this gap is determinant for the Curie temperature $T_C$  that  marks  the  quenching of ferromagnetism. $T_C$ also depends on the exchange coupling between spins, the magnitude of the spins, and the number of nearest neighbours\cite{Majlis_book_2007}. In the case of localized spins at the atomic sites, e.g., in insulating systems, one typically uses a Heisenberg-like spin Hamiltonian that includes, at least, nearest neighbours isotropic exchange interactions between sites, as well as the single-ion anisotropy at each site and, additionally, anisotropic exchange or even antisymmetric exchange\cite{Abragam_Bleaney_book_1970}. Single-ion anisotropy is determined by the crystal field around the magnetic atoms and the strength of spin-orbit interaction, this latter being also relevant for the anisotropic exchange. However, explaining the particular type of exchange interactions for each and every system is far from trivial and different models have been proposed\cite{elton_2022}. 
%

From the experimental side, there are a bunch of experimental techniques that can be used to explore the magnetic order of thin films with magnetic anisotropy and, thus, the signatures of both single and multi-magnon processes. This includes X-ray magnetic circular dichroism (XMCD)~\cite{Stohr_jesrp_1995} and its depth-resolved variants~\cite{Amemiya_Kitagawa_apl_2004,Sakamaki_Amemiya_rsi_2017}, neutron scattering, which has the additional advantage of direct access to the dispersion relation on the whole wave vector space~\cite{Majkrzak_2005,Callori_Saerbeck_ssp_2020}, magnetometry using SQUID~\cite{Vettoliere_Silvestrini_book_2023}, Raman spectroscopy~\cite{Fleury_Guggenheim_prl_1970,Devereaux_Hack_rmp_2007,Hien_Thi_jrs_2010,Zhang_Wu_nanoll_2020}, ferromagnetic resonance (FMR)~\cite{Kittel_pr_1948,Beaujour_Ravelosona_prb_2009,Usov_jmmm_2019}  or local techniques, such as spin-polarized scanning tunneling microscopy~\cite{Wiesendanger_revmod_2009} and inelastic tunneling spectroscopy~\cite{Klein_2018, Gao_2008,Balashov_2006}. Interestingly, only recently the effect of magnons in atomically thin layers has been accessible experimentally through Raman scattering~\cite{Cenker_Huang_nature_2021}, allowing for instance to determine the optical selection rules established by the interplay between crystal symmetry, layer number, and magnetic states in CrI$_3$.

Nowadays, the use of first-principles calculations permits obtaining accurate total energy values for different spin configurations of a given magnetic system. A proper choice of these configurations with different spin orientations allows fitting parameters defined in the spin Hamiltonian by mapping total energy differences, although with some limitations \cite{Sabani_2020}.
%
In this work, we have considered four two-dimensional systems of different kinds: 1) single septuple layer of the magnetic topological insulator MnBi$_2$Te$_4$, 2) single triple layer of a Mott-Hund´s magnetic insulator CrI$_3$, 3) 2D metal-organic coordination network Fe-DCA on the Au(111) substrate, and 4) monolayer of Co on the heavy metal substrate Pt(111). The  four  systems  have been shown~\cite{Otrokov_2019,Otrokov_Nat_2019,Huang_2017,Lobo_2023,Zimmermann_Bilhmayer_prb_2019} to present ferromagnetic order at finite temperatures and, therefore, are interesting to consider. Our aim here is to understand the nature and origin of the anisotropy \cite{Lado_Rossier_2dmat_2017,soriano_2020}, as well as the spin wave excitation spectrum and Curie temperatures. In order to achieve a qualitative understanding, we use an auxiliary toy model of finite ferromagnetic spin chains with exchange and anisotropy parameters extracted from the previous mapping of the spin Hamiltonian.

\section{Density Functional Theory calculations and mapping to spin Hamiltonian\label{DFT-methods}}
In this section we briefly describe the way density functional theory (DFT) calculations have been performed for the four magnetic systems of interest, as well as the extraction of the exchange and anisotropy parameters that are later on used to model the spin wave excitations. 


DFT calculations were performed using the projector augmented-wave method \cite{Blochl.prb1994}, implemented in the VASP code \cite{vasp1,vasp2,vasp3}. The generalized gradient approximation (GGA) was employed to describe the exchange-correlation energy \cite{Perdew.prl1996}. To take into account the strongly localized character of the $3d$-states (except those of Co in the Co/Pt system, which is metallic), we resorted to the GGA + $U$ approach \cite{Anisimov1991}. The $U_{\rm eff}=U-{\cal J}$ \cite{Dudarev.prb1998} parameter was chosen to be equal to 2, 6, and 5.34 eV for Cr, Fe, and Mn $3d$-states, respectively, following the literature \cite{Lado_Rossier_2dmat_2017, Lobo_2023, Otrokov_Nat_2019}. For each system of interest, the atomic coordinates were relaxed using a force tolerance criterion for convergence
of 0.01 eV/{\AA}.

\begin{figure}
    \centering
    \includegraphics[width=0.45\textwidth]{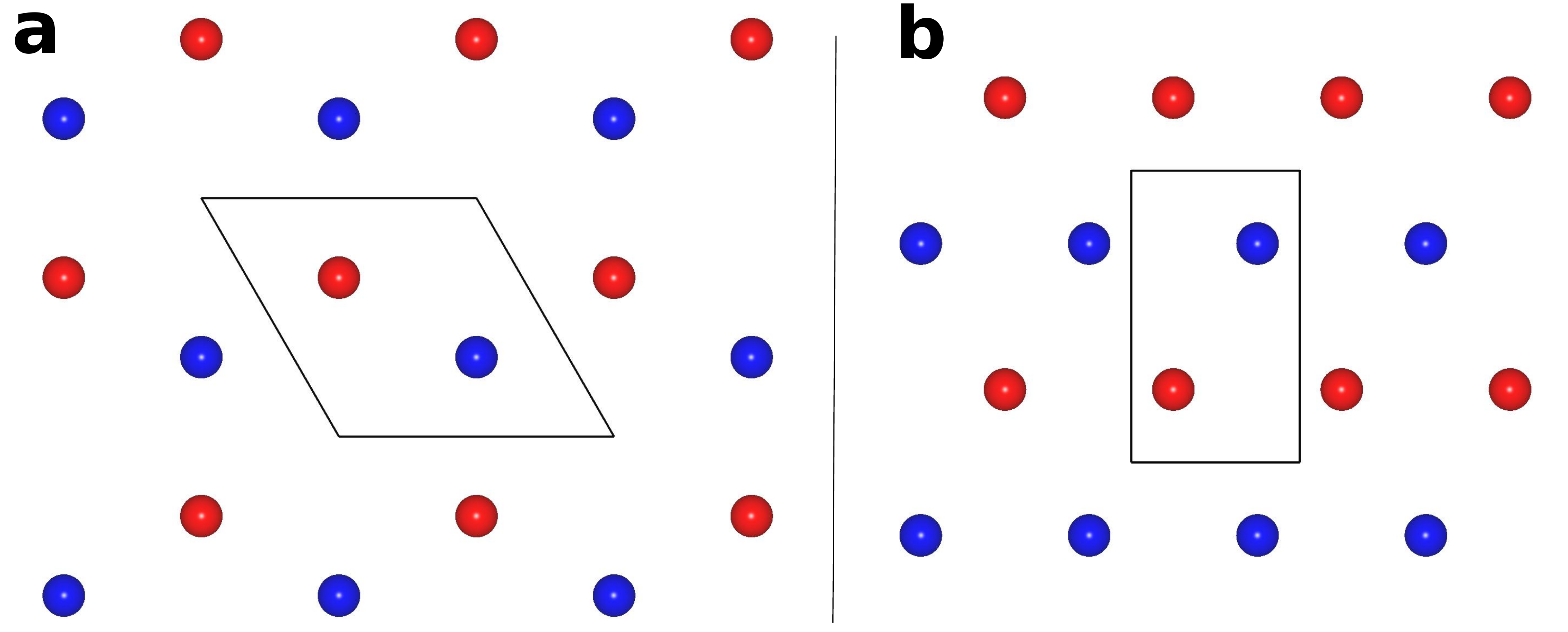}
    \caption{Top view of the two-dimensional structures of the systems considered (only magnetic atoms are drawn) showing the unit cells used in the calculations for the $(1 \times {1})$ honeycomb (a) and $(1 \times \sqrt{3})$ hexagonal (b) layer cases, both containing two atoms in the unit cell. The blue and red colors represent opposite spins in an antiferromagnetic configuration to illustrate the number of first neighbours with aligned and anti-aligned spins for each lattice: three anti-aligned spins in the honeycomb (a), while four anti-aligned and two aligned spins in the hexagonal (b). Notice that in a ferromagnetic configuration of spins the number of first neighbours with aligned spins coincides with the coordination, i.e, three in the honeycomb and six in the hexagonal lattice.}
    \label{fig0}
\end{figure}

The calculations were performed in a model of repeating films separated by a vacuum gap of a minimum of 10~\AA. In the Co/Pt and MnBi$_2$Te$_4$ systems [CrI$_3$ and Fe-DCA/Au(111) systems], in which the $3d$-atom layer has a hexagonal [honeycomb] lattice, the $(1 \times \sqrt{3})$ [$(1 \times {1})$] \emph{magnetic} periodicity was used, so that the unit cell contains two \emph{magnetic} atoms in each case (see Fig. \ref{fig0}).

The values of the nearest-neighbor Heisenberg exchange coupling constants $J$, anisotropic exchange $\lambda$ [$(J^z-J)$ in Table \ref{table-Tc}], and single-ion anisotropy $D$ parameters [see equation (\ref{Hgen}) below where the spin Hamiltonian is explicitly given] were obtained via accurate static total-energy calculations, performed for the optimized crystal structures. 
While $J$ is obtained via a scalar relativistic DFT calculation of ferromagnetic and antiferromagnetic configurations, determining $\lambda$ and $D$ requires (i) inclusion of spin-orbit coupling and (ii) consideration of the latter two magnetic configurations for both in-plane and out-of-plane moment directions. In all of these static total-energy calculations we used a strict convergence criterion of 10$^{-8}$ eV, the $\overline \Gamma$-centered $k$-points sampling of the 2D Brillouin zone, and the tetrahedron integration method with Bl\"{o}chl corrections.
The specific Brillouin zone samplings are given in the Supplementary Information.

For the systems with a honeycomb lattice (Cr and Fe-containing systems), there are two magnetic atoms in the surface unit cell, and each atom is coordinated with three nearest neighbors (Fig. \ref{fig0}a). These three nearest neighbours can have ferromagnetic $-3JS^2$ or antiferromagnetic $3JS^2$ coupling between them, which translates into a total energy difference of $6JS^2$.  For Co/Pt(111) and MnBi$_2$Te$_4$, the coordination of Co and Mn atoms is six and we have to use a $(1 \times \sqrt{3})$ unit cell that contains two magnetic atoms. For the cell shown in Fig.~\ref{fig0}b, the ferromagnetic coupling includes the six equal $-JS^2$ contributions, i.e., a total of $-6JS^2$, while the antiferromagnetic coupling includes two spins coupled ferromagnetically and four spins coupled antiferromagnetically, i.e., a total of $2JS^2$, which translates into a total energy difference of $8JS^2$. This number of aligned or antialigned spin pairs counting is the same for the anisotropic exchange contribution, while the single-ion anisotropy term includes only two magnetic atoms per cell. 
%

\begin{table}[]
\resizebox{\columnwidth}{!}{\begin{tabular}{|l|r|r|r|rr|}
\hline
\multicolumn{1}{|c|}{\multirow{2}{*}{System}} & \multicolumn{1}{c|}{\multirow{2}{*}{$J$ (meV)}} & \multicolumn{1}{c|}{\multirow{2}{*}{$D$ (meV)}} & \multicolumn{1}{c|}{\multirow{2}{*}{$\lambda$ (meV)}} & \multicolumn{2}{c|}{$T_C$}                  \\ \cline{5-6} 
\multicolumn{1}{|c|}{}                        & \multicolumn{1}{c|}{}                           & \multicolumn{1}{c|}{}                           & \multicolumn{1}{c|}{}                                 & \multicolumn{1}{c|}{$k_BT_C$(meV)} & \multicolumn{1}{c|}{$T_C$(K)} \\ \hline
Co/Pt ($S=1$)                                  & 55.0                                            & 1.20                                           & -0.11                                                 & \multicolumn{1}{r|}{26.9}     & 312                        \\ \hline
CrI$_3$ ($S=3/2$)                             & 3.75                                            & 0.039                                          & +0.16                                                 & \multicolumn{1}{r|}{3.73}     & 43.2                       \\ \hline
Fe-DCA\misha{/Au} ($S=2$)                                & 0.72                                            & 0.19                                           & -0.021                                                & \multicolumn{1}{r|}{1.50}     & 17.4                       \\ \hline
MnBi$_2$Te$_4$ ($S=5/2$)                                 & 0.22                                            & 0.056                                          & -0.0075                                               & \multicolumn{1}{r|}{1.27}     & 14.76                       \\ \hline
\end{tabular}}
\caption{ 2D systems under study with the spin $S$ of each magnetic atom (within brackets), the corresponding values of exchange couplings $J$, axial anisotropy $D$ and anisotropic exchange $\lambda=J_z-J$ as extracted from DFT according to the mapping with Hamiltonian (\ref{Hgen}).  $T_C$ was calculated according to the expressions provided in Ref.~\cite{Torelli_Olsen_2dmat_2019}.
}
\label{table-Tc}
\end{table}

The results obtained for the four systems under study are given in Table \ref{table-Tc}. In all cases, the largest energy scale 
corresponds to the isotropic exchange strength, which can be as large as tens of meV in the Co/Pt case but of the order of meV, or even less, in the three other cases. As shown in the next section, the isotropic exchange is crucial to determine the Curie temperature. The anisotropic exchange $\lambda$ is, at least, two orders of magnitude smaller than the isotropic exchange $J$.
The single-ion anisotropy $D$ is larger than the anisotropic exchange, except for CrI$_3$. The largest value of $D/J = 0.26$ corresponds to the case of MnBi$_2$Te$_4$. These variations in the anisotropic exchange and single-ion anisotropy, relative to the isotropic exchange, have consequences in the spin wave excitation spectrum that are discussed in section \ref{results_sw}: they will induce an energy gap in the spin waves excitation spectra.
%
%

\section{Critical temperatures}
%
The magnetic susceptibility of a ferromagnet obeys Curie-Weiss law where, above a critical temperature $T_C$, it behaves essentially as a paramagnet with an enhanced susceptibility~\cite{Yosida}, while below it, the material can keep a finite magnetization in the absence of applied field. Thus, it is important to estimate reasonable $T_C$ values for the different systems.  Torelli and Olsen~\cite{Torelli_Olsen_2dmat_2019} proposed a simple parametric dependence of $T_C$ on the model parameters for different two-dimensional materials based on fitting to classical Monte Carlo simulations. The proposed $T_C$ reproduces both the critical temperature in the large magnetic anisotropy limit, and in the low anisotropy limit where the commonly used random phase approximation (RPA)~\cite{Yosida} fails. Table \ref{table-Tc} shows the critical temperatures extracted for the four materials studied in Sec.~\ref{DFT-methods}. Notice that since $D/J,\;\lambda/J\ll 1$ in all cases, the value of the critical temperature is essentially determined by the value of the isotropic exchange $J$, the number of nearest neighbours and the magnitude of the spin magnetic moment $S$ \cite{Torelli_Olsen_2dmat_2019}. However, it requires finite values of the anisotropy to differ from $T_C=0$. Incidentally, these calculated $T_C$  are in reasonable agreement with available data or estimated values by other methods~\cite{Cenker_Huang_nature_2021,Huang_2017,Otrokov_2019,Lobo_2023,Zimmermann_Bilhmayer_prb_2019}.
%
%

%
\begin{table}[t!]
\begin{center}
\begin{tabular}{|c|c|c|c|}
\hline
System & $S$ & $D/J$ & $(J^z-J)/J$ \\ \hline
A & 1 & 0.02 & -0.003 \\ \hline 
B & 3/2 & 0.01 & 0.04 \\ \hline 
C  & 2 & 0.25 & -0.03 \\ \hline  
D & 5/2 & 0.26 & -0.03 \\ \hline 
\end{tabular}
\end{center}
\caption{Values of the spin $S$, relative local magnetic anisotropy $D/J$ and anisotropic exchange $(J^z-J)/J$ for four different spin rings whose parameters resemble those of Co/Pt, CrI$_3$, Fe-DCA/Au and MnBi$_2$Te$_4$, respectively. For simplicity, we have denoted $J^\perp =J$. }
\label{table-param}
\end{table}

\section{Spin wave excitations in finite ferromagnetic spin arrays}
We consider a system of $N$ interacting spins described by the following spin Hamiltonian:
\beqa
H&=&-\frac{1}{2}\sum_{\langle i,j\rangle}\left[J^z_{ij}
S_i^z S_j^z+2J_{ij}^\perp\left( S_i^+S_j^- +S_i^- S_j^+\right)\right]
\crcr
&& \qquad -D\sum_{i}\left(S_i^z\right)^2+g\mu_B B \sum_i S^z_i,
\label{Hgen}
\eeqa
where $J^z_{ij}\ge 0$ represents the exchange coupling of the longitudinal component $z$ between spins $i$ and $j$, $J^\perp_{ij}$ is the exchange coupling on the perpendicular plane, and thus, $\lambda_{ij}=J^z_{ij}-J^\perp_{ij}$ 
corresponds to the anisotropic exchange between the pair of spins $(i,j)$. 
In addition, $D$ is the uniaxial magnetic anisotropy ($D>0$ conforms with an easy axis magnetic anisotropy). The operators $S_j^z$ and  $S_j^\pm$ correspond to the $z$-component of the $j$-spin operator and the corresponding ladder operators $S_j^\pm=S_j^x\pm i S_j^y$.
Here the sum over indices $\langle i,j\rangle$ denotes the sum over the first neighbors. 
This Hamiltonian can represent either a one-dimensional or a multidimensional system. 
For one-dimensional rings, we impose periodic boundary conditions where ${\bf S}_1={\bf S}_N$.

For the description of the spin waves, it is convenient to introduce the Fourier transforms of the exchange interaction
\beq
J^{z/\perp}({\bf q})=\sum_{j} J_{kj}^{z/\perp}e^{i{\bf q}\cdot\left({\bf R}_j-{\bf R}_k\right)},
\eeq
where we have assumed that the local ${\bf S}_j$-spin is at the ${\bf R}_j$ position, and the dimensionless parameter
\beq
\xi =J^\perp({\bf q})/J^z({\bf q})=J_{k,j}^\perp/J_{k,j}^z,\quad \forall j,k.
\nonumber
\eeq

Notice that the periodicity of the lattice implies that $J^{z/\perp}({\bf q})$ does not depend on the $k$-lattice site.
For simplicity, we assume that all spins are of equal magnitude $S\ge 1/2$ and a uniform exchange interaction between first-neighbors, i.e., $J_{kj}^{z/\perp}=J^{z/\perp}$ if $j$ and $k$ are first-neighbors and zero otherwise.

The ground state $|G\rangle$ of Hamiltonian (\ref{Hgen}) for $J^{z},\; J^{\perp}>0$ corresponds to the Weiss state that can be written as $|G\rangle= \prod_{i=1}^{N} |-S\rangle$, being $S$ the total number of spins in the lattice.\footnote{The degeneracy of the ground state level could be broken by an infinitesimal applied field $B_z>0$, in which case the $S_z=-S$ projection is favored.} For finite-size systems, one can in principle solve Hamiltonian (\ref{Hgen}) and find the discrete energy levels $\epsilon_n$ and the corresponding eigenvectors $|\epsilon_n\rangle$. These states can be written in terms of the spin configurations $|\alpha\rangle \equiv |m_1,m_2,\dots,m_N\rangle$, where $m_i$ is the spin quantum number associated with $S^z_i$. Then, we can write the eigenvectors of $H$ as
$|\epsilon_n\rangle= \sum_\alpha \langle \alpha|\epsilon_n\rangle |\alpha\rangle$.

\subsection{Single-spin excitation}%
Single magnons are delocalized spin waves whose total spin differs in one unit of angular momentum with respect to the ground state of the magnet. In our finite-size system described by Eq. (\ref{Hgen}), 
we define the normalized local excitation of the $\ell$ spin of the Weiss state as
\beq
|\ell \rangle=\frac{1}{\sqrt{2S}}S_\ell^+|G\rangle.
\label{SW-l}
\eeq
 Here we have included the normalization factors similarly to what is done in usual spin wave theory (see supporting information for further details). A finite-size spin wave-like state containing a single magnon has the form
\beq
|{\bf q}\rangle=\frac{1}{\sqrt{N}}\sum_{\ell=1}^N e^{i{\bf q}\cdot {\bf R}_\ell}|\ell\rangle,
\label{SW-q}
\eeq
with ${\bf R}_{ \ell}$ the position of the $\ell$-spin. Notice that, for a closed one-dimensional chain, the momentum $q$ is quantized due to the periodic boundary conditions (see supplementary material). 
 Thus, we can define the overlap of the eigenvectors of our spin Hamiltonian with the $|{\bf q}\rangle$ spin wave as
\beq
 \psi_{\bf q}^n\equiv \langle {\bf q}|\epsilon_n\rangle=
 \frac{1}{\sqrt{N}}\sum_{\ell}e^{-i{\bf q}\cdot {\bf R}_{\ell}}\sum_{\alpha}\langle \ell|\alpha\rangle\langle\alpha|\epsilon_n\rangle,
\eeq
where one finds that
\beqa
\langle \ell |\alpha\rangle &\equiv& \frac{1}{\sqrt{2S}}\langle G|S_{\ell}^-|m_1,\dots, m_N\rangle=
\delta_{m_1,-S}\dots \delta_{m_\ell-1,-S}
\crcr
&\times& \delta_{m_\ell,-S+1}\delta_{m_\ell+1,-S}\dots\delta_{m_N,-S}.
\eeqa

The excitation energy of a single spin wave can be found from semiclassical arguments and it reads as~\cite{Majlis_book_2007}:
\beqa
E_{SW}({\bf q})&=&S\left(J^\perp({\bf 0})/\xi-J^\perp({\bf q})\right)+D\left(2S-1\right)+g\mu_B B,
\crcr
&&
\label{esmagnon}
\eeqa
where $\xi =J^\perp({\bf q})/J^z({\bf q})$.
Hence, the spin-wave spectrum presents an energy gap of magnitude
$S\left(J^\perp({\bf 0})/\xi-J^\perp({\bf 0})\right)+D\left(2S-1\right)$ at zero magnetic field.
In finite-size systems, the momentum ${\bf q}$ is quantized into a set of discrete values $\left\{{\bf q}_i\right\}$ and, thus, for each energy level $\epsilon_n$ there will be a given number ${\cal N}_n$ of states with overlaps $\psi_{{\bf q}_i}^n$ above a given threshold $\varepsilon$. Moreover, when finite-size effects are negligible, 
not only does the lowest energy $\epsilon_n$ reproduce the spin-wave dispersion $E_{\rm SW}({\bf q})$ perfectly~\cite{Gauyacq_Lorente_prb_2011} but, as we shall see below, it leads to a single state with non-negligible overlap for $E_{\rm SW}({\bf q}_i)=\epsilon_n$.

\subsection{Two-spin excitation\label{TMexc}}
%
Two-magnon excitations of a ferromagnet correspond to spin waves differing by two units of angular momentum from the totally aligned states, a set of states that comprises an invariant subspace under the dynamical motion defined by the Heisenberg Hamiltonian~\cite{Bethe_zfp_1931}. 
These two-magnon excitations depend on the momentum of the pair of magnons ${\bf k}_1$ and ${\bf k}_2$ or, alternatively, on the total momentum ${\bf Q}$ and the difference ${\bf q}$, where ${\bf k_1}={\bf Q}/2+{\bf q}$ and ${\bf k_2}={\bf Q}/2-{\bf q}$. 
It embraces a continuum of states corresponding to two non-interacting magnons. In addition, there may be up to two branches below the continuum that correspond to two-magnon bound states resulting from the magnon-magnon interaction that confines both magnons together~\cite{Bethe_zfp_1931}. 

 %
%
\begin{figure*}[t!]
    \centering
    \includegraphics[width=0.9\textwidth]{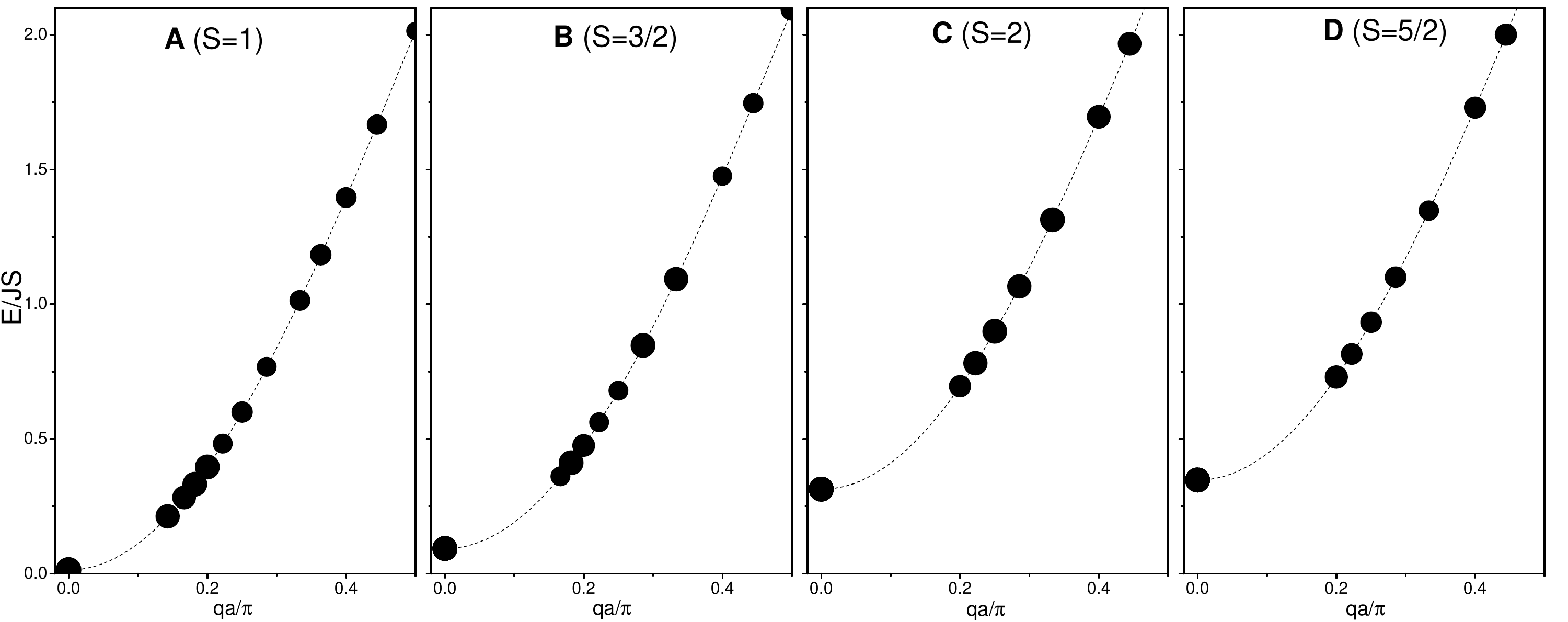}
    \caption{Single magnon excitations of finite-size rings. Single-magnon energies (dashed lines) as calculated from $E_{SW}(q)$ and from the diagonalization of the finite-size Hamiltonian (\ref{Hgen}) for a one-dimensional ring (black circles) corresponding to A, B, C, and D systems in Table \ref{table-param}. Each non-zero momentum $q=2\pi n/(Na)$ is associated with a single chain of size $N$. The symbol's sizes are proportional to the weights $\psi_{q}^n$ for the single magnon, and only states with projections $\psi_{q}^n$ above a threshold value of $\epsilon=0.1$ have been depicted. }
    \label{fig:1}
\end{figure*}

Two-magnon excitations are essential to understand the low-temperature Raman scattering $({\bf Q}={\bf 0}$) where they manifest as resonances close to the bound states energies~\cite{Loly_prb_1976}, the enhanced relaxation of uniform modes observed in FMR~\cite{Beaujour_Ravelosona_prb_2009}, or the strongly temperature-dependent peaks in neutron scattering~\cite{Cowley_Buyers_prl_1969,Huberman_Coldea_prb_2005,Korner_Lenz_prb_2013}. Moreover, exchange Hamiltonians like (\ref{Hgen}) display two-magnon bound states bellow a two-magnon continuum~\cite{Hanus_prl_1963,Wortis_pr_1963}.
 These multiple-magnon bound states can be detected in the resonant spectra~\cite{Date_Motokawa_prl_1966,Torrance_Tinkham_pr_1969}
  or in electron spin resonance (ESR) experiments~\cite{Hoogerbeets_Duyneveldt_jpc_1984}, with intensities that decrease exponentially upon lowering the temperature or increasing the multiplicity of the magnon mode. However, due to the dominant statistical weight of the two-magnon continuum in thermodynamic measurements, the common forbidden transition character of the ${\bf Q}={\bf 0}$ point explored in most spectroscopic techniques~\cite{Hoogerbeets_Duyneveldt_jpc_1984},
  the inherently weak intensity of the two-magnon cross-section
  makes the observation of these bound states quite challenging. Nevertheless, they can play an important role in the entanglement and non-equilibrium dynamics as explicitly in trapped ions quantum simulators~\cite{Fukuhara_Schauss_nature_2013,Kranzl_Birnkammer_prx_2023}.

The problem of the two magnons has been solved exactly by Tonegawa~\cite{Tonegawa_ptp_1970} in one and two dimensions (the main results can be found in the supplementary material). 
He found that there are two branches of the energy of the two-magnon bound state $E_{2M}({\bf Q})$. He observed that below a certain threshold momentum $Q_{th}$, there is a pair of complex conjugate solutions and one real solution.  This indicates that the corresponding bound-state branch is no longer stable and the bound-state dispersion crosses the continuum.
Moreover, for certain regions of the parameter space, negative values of the (excitation) energy of the two-magnon-bound state can be obtained. This reflects that one-dimensional ferromagnets may show unstable fully aligned states along the quantization axis $z$ even for parameters for which the single magnon excitation energy is positive.  
 In addition, it is possible to find single-magnon energies $E_{SW}({\bf q})$ above the two-magnon bound states $E_{2M}(Q)$.

In particular, he found that the amplitude $\psi_Q(\ell,\ell')$ for finding two spin-deviations at the $\ell$-th and $\ell'$-th sites decays quite fast with $|\ell-\ell'|$, with $\psi_Q(\ell,\ell')=\delta_{\ell,\ell'}$ at the Brillouin zone boundary for one of the two energy branches, the so-called {\em Ising type}, while $\psi_Q(\ell,\ell')=\delta_{\ell',\ell+1}$ for the second one, called as {\em Bethe type}.
 For an arbitrary value of ${\bf Q}$, one of the two magnon bound states has a larger amplitude on nearest neighbors while the second one has a larger amplitude on the same site, but they can not be classified rigorously into either type~\cite{Tonegawa_ptp_1970}. In the same way, the amplitudes $\psi_Q(\ell,\ell')$ decay much slower with $|\ell-\ell'|$ as we move away from the first Brillouin zone boundaries.

Let us formulate the problem using the ideas of spin-wave theory.
The normalized two-spin excitation corresponding to sites $\ell$ and $\ell'$ is defined as 
\beq
|\ell,\ell' \rangle=\chi_{\ell\ell'}^S S_\ell^+ S_{\ell'}^+|G\rangle,
\eeq
where $\chi_{\ell\ell'}^S=1/2S$ if $\ell\ne \ell'$ and $\chi_{\ell\ell}^S=1/\sqrt{4S(2S-1)}$. 
Similarly to the single spin excitations, we can define a two-magnon wave function as
\beqa
|{\bf Q},{\bf q}\rangle&=& \frac{1}{\sqrt{N}}\sum_{\ell,\ell'}
e^{i{\bf Q}\cdot \left({\bf R}_\ell+{\bf R}_{\ell'}\right) } e^{i{\bf q}\cdot\left({\bf R}_\ell-{\bf R}_{\ell'}   \right)}b_{|\ell-\ell'|}|\ell,\ell'\rangle,
\crcr
&&
\label{TMwave}
\eeqa
where the normalization condition imposes that $\sum_{\ell\ell'}\left|b_{|\ell-\ell'|}\right|=N$. We now define the overlaps of the eigenvectors $|\epsilon_n\rangle$ with the two-magnon waves 
\beqa
\psi_{{\bf Q},{\bf q}}^{\rm Bethe}(n)&\equiv & 
\frac{1}{\sqrt{N}}\sum_{\ell}\sum_{n.n.,\ell}e^{-i{\bf Q}\cdot( {\bf R}_{\ell}+{\bf R}_{n.n.,\ell})}
\crcr
&&\quad \times e^{-i{\bf q}\cdot( {\bf R}_{\ell}-{\bf R}_{n.n.,\ell})}\sum_{\alpha}\langle \ell,n.n.|\alpha\rangle\langle\alpha|\epsilon_n \rangle
\crcr
\psi_{{\bf Q}}^{\rm Ising}(n)&\equiv & 
\frac{1}{\sqrt{N}}\sum_{\ell}e^{-2i{\bf Q}\cdot {\bf R}_{\ell}}\sum_{\alpha}\langle \ell,\ell|\alpha\rangle\langle\alpha|\epsilon_n \rangle,
\label{TM_overl}
\eeqa
 where $n.n.,\ell$ indicates that the sum is realized over the nearest neighbors of spin $\ell$. 
 Markedly, while for the Ising-type the two-magnon is fully defined by a total momentum ${\bf Q}$, the Bethe-type two-magnon state also depends on the momentum difference ${\bf q}$ and so does also the overlap $\psi_{{\bf Q},{\bf q}}^{\rm Bethe}(n)$. These overlaps contain, in addition to the two-magnon bound state contribution, a portion of two unbound magnons that, in average, occupy either the same or neighboring sites for each type, respectively~\cite{Kranzl_Birnkammer_prx_2023}.  

 %
As it happens for the single magnon, for each energy level $\epsilon_n$, there will be a finite number of states with overlaps $\psi_{\bf Q,q}^n$ above a given threshold.

 \subsection{Results for a one-dimensional toy model\label{results_sw}}
%
To further extend our understanding of the spin waves and their connection with the magnetic anisotropy,
we will analyze the energy spectrum of a closed chain of anisotropic spins. 
We stress that, from the practical point of view, the main effect of the change in dimensionality is to extend the sum over the first neighbors.
For the ring, the momentum $q$ of the spin waves is quantized, i.e., $q_n= 2\pi n/(aN),\; n=0,\dots, N-1$, where $a$ is the distance between spins in the chain. Here, we have considered four scenarios that mimic the properties of the systems studied in Sec. \ref{DFT-methods}. The model parameters of each of these systems are summarized in Table \ref{table-param}. The four cases A-D correspond to situations of increasing anisotropy, starting from the almost isotropic Heisenberg-like $S=1$ chain in A and finishing with the largest anisotropy-induced gap for case D.

%
%
\begin{figure*}[t!]
    \centering
    \includegraphics[width=0.9\textwidth]{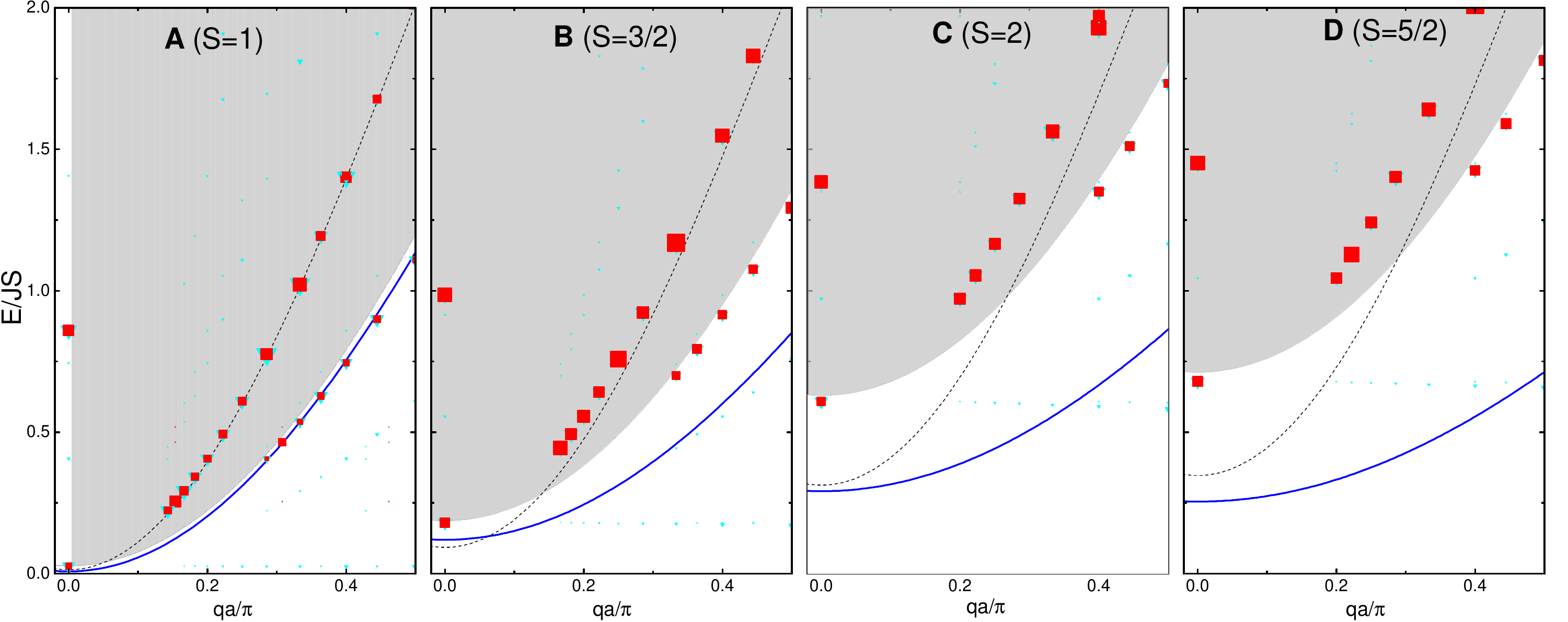}
    \caption{Two-magnon spectra for systems A-D. The shaded area represents the two-magnon continuum $E_{2SW}$, while the blue solid line corresponds to the two-magnon bound state energy $E_{2M}$. The symbols represent the 
 energies $\epsilon_n$ from the exact diagonalization of Hamiltonian (\ref{Hgen}) of $N$ spins and the corresponding weights $\psi_{Q,0}^{\rm Bethe}(n)$ for the Bethe-type (cyan triangles), and $\psi_{q}^{\rm Ising}(n)$ for the Ising-type (red squares) two-magnon spin waves (only weights above 0.1 have been represented). The symbol's sizes are proportional to the weights. 
    The single magnon solution $E_{SW}(Q)$ (dashed line) has been added as a reference.  Each non-zero momentum $q = 2\pi n/(Na)$ is associated with a single chain size $N$.   }
    \label{fig2}
\end{figure*}

%
%
%
Figure \ref{fig:1} summarizes the main results of the spin waves in close rings with the parameters of Table \ref{table-param}. The single spin wave solution $E_{SW}(q)$ has been plotted as a reference for the four cases.
%
Case A represents the closest case to the isotropic Heisenberg spin chain.  As observed, it does not show any apparent gap in the spectrum of spin waves. The presence of a local anisotropy $D\ne 0$ or an anisotropic exchange induces a finite energy gap at $k=0$ of magnitude $E_{SW}^{\rm gap}=2SJ^\perp \left(1/\xi-1\right)+D\left(2S-1\right)$.  This is clearly observed in cases C and D (notice that the gap has been scaled by $JS$).

Let us now analyze what happens in finite-size rings and, in particular, to the overlaps with the single-magnon. 
In all cases, the single-magnon excitations overlap exactly with the lattice solution due to the preserved translational symmetry of the ring. Hence, the only effect of the finite size is the quantization of the momentum $q$. Interestingly, only the energy level with $\epsilon_n-\epsilon_0=E_{\rm SW}^{\rm ring}(q)$ shows a significant overlap with the single magnon spin wave, with $\epsilon_0$ the ground state energy. This is depicted in Fig.~\ref{fig:1} by the size of the corresponding symbols. 

%
%
As explained in Sec.~\ref{TMexc}, two-magnon excitations can be relevant in the dynamics of these spin systems, in particular, in the low-temperature limit. With this idea in mind, we have also explored the two-magnon excitations of the above systems, see Fig. \ref{fig2}. We have represented the two-magnon continuum by the shaded area together with the energy of the two-magnon bound states (see Supplementary Material for detailed expressions of continuum $E_{2SW}(Q,q)$ and bound states energies $E_{2M}(Q)$).
For the parameters in A-D, only one of the solutions corresponding to the bound states is positive and real.
Notice that the two-magnon bound state energies constitute a lower bound to the energy of the continuum, and though the bound state energies approach the bottom of the continuum as $Q\to 0$, it may or may not merge with it~\cite{Tonegawa_ptp_1970,Sharma_Lee_prb_2022}.

Interestingly, an energy gap can also be opened on the two-magnon-bound states' excitations,  and it can lie either above (as in B) or below (D) the single-magnon excitation,
a situation typically found for large enough anisotropy $D$ and where a large impact of two-magnon excitations is expected~\cite{Tonegawa_ptp_1970}. Indeed, Tonegawa~\cite{Tonegawa_ptp_1970} argued that this situation indicates that the states in which all the spins aligned along the $z$-axis are unstable.

Let us focus now on the projection of the energy eigenstates of (\ref{Hgen}) on the two-magnon states for finite rings.  To facilitate the interpretation,  only the $N=10$ energies $\epsilon_n$ have been depicted for $q=0$, the only case where the different chain sizes correspond to the same momentum.
Importantly, the long wavelength limit maintains the single magnon as the lowest energy excitation.
%
As advanced in Sec.~\ref{TMexc}, the overlaps defined in Eq. (\ref{TM_overl}) contain a contribution of single-magnon type clearly reflected in Fig.~\ref{fig2}.
For almost isotropic rings (case A), there is a quite large overlap with the two-magnon-bound state for arbitrary  momentum, which shows both Ising and Bethe-like character in the low-$k$ region.

The situation changes quite drastically when anisotropy is included. 
First, the quantized two-magnon excitation deviates appreciably from the semiclassical curve of the infinite lattice. In particular, the single magnon-like contribution lies above the single-magnon semiclassical solution, well within the two-magnon continuum and having a prominent Ising-type character, see cases B-D.
Second, the two-magnon bound states, which for the isotropic case almost overlap with the semiclassical solution (see case A), now also move upwards in energy (cases B-D) when anisotropy increases. Indeed, for large anisotropy the long wavelength lowest energy excitation in the semiclassical limit corresponds to two-magnon excitations and, therefore, it represents a qualitative difference with respect to the isotropic case.
Finally, contrary to the single magnon case where there is a single eigenstate of $H$ that does overlap with the spin wave and, hence, has a well-defined character, now a large number of eigenstates with both Ising- and Bethe-type characters display finite but small overlap with the two-magnon sates (\ref{TMwave}).

\section{Summary and conclusions}
Here, we have estimated the Curie temperatures of different two-dimensional materials based on the simple phenomenological expressions provided by Torelli and Olsen~\cite{Torelli_Olsen_2dmat_2019}. In so doing, the isotropic and anisotropic exchange coupling constants, as well as the single-ion anisotropy, are obtained by mapping the total energy differences obtained from DFT calculations for different spin configurations with a spin Hamiltonian model.

Later on, in order to explore the low-energy magnetic excitations in systems with magnetic anisotropy, we use a simple finite-size periodic chain model of spin Hamiltonian with the previously calculated exchange coupling constants and anisotropy. The corresponding eingenstates are found by exact diagonalization. Their projection onto single-magnon and two-magnon states reveals important changes in the spin wave excitation spectrum for large values of the magnetic anisotropy, both single-ion and anisotropic exchange. We not only reproduce the well-known opening of a gap in the single-magnon excitations, but also find the importance of two-magnon excitations in the low-energy spin wave excitation spectrum.
In addition, two-magnon excitations of different kinds, including two-magnon bound states~\cite{PhysRevB.2.772}, are found. 
These results suggest that some of the two-dimensional materials considered in this work, particularly those with large magnetic anisotropy, may present a lower two-magnon energy gap compared to the usual single-magnon excitation. Therefore, we
speculate that, in systems with large magnetic anisotropy, multi-magnon processes can play an important role in determining the low-energy magnetic excitations, although the intensity of the corresponding signal is expected to be weaker \cite{Elnaggar2023} and, thus, are relevant in the interpretation of the low-temperature properties of these two-dimensional ferromagnets. Its confirmation, of course, requires the use of precise and sensitive techniques, like Raman scattering \cite{Cenker_Huang_nature_2021} or ferromagnetic resonance \cite{Kittel_pr_1948,Beaujour_Ravelosona_prb_2009,Usov_jmmm_2019}.

\begin{acknowledgments}
We are grateful to N. Lorente and Leonid M. Sandratskii for fruitful discussions.
This work was supported by MCIN/ AEI /10.13039/ 501100011033/  (Grants
PID2022-138269NB-I00, 
PID2019-103910GB-I00, PID2022-137685NB-I00, 
and PID2022-138210NB-I00
) and FEDER ``Una manera de hacer Europa". 
 We also acknowledge the support by the University of the Basque Country (Grant no. IT1527-22) 
 \end{acknowledgments}



\clearpage
\newpage
\section*{Supporting Material\label{SOM}}

\section*{Supercell structure and Brillouin zones in DFT calculations}
\subsubsection*{Co/Pt.} For Co monolayer on Pt(111), the substrate contained five atomic layers. The two lowermost layers of Pt were fixed upon structure optimization, while all other layers in the cell were allowed to relax along the out-of-plane $z$-direction. The optimized Pt bulk lattice parameter $a_0=2.813$ \AA\, was used to define the in-plane $(1 \times \sqrt{3})$ cell of  rectangular shape. The Co monolayer was expanded to match the Pt lattice constant. 
The static total-energy calculations were performed using the $43\times 25\times 1$ $k$-mesh.

\subsubsection*{CrI$_3$.} The optimized bulk in-plane lattice constant $a_0=7.1$ \AA\, was used for the free-standing CrI$_3$ trilayer. This corresponds the Cr-Cr separation of 4.1 \AA. The Cr-I interlayer distances were relaxed. 
The static total energy calculations were performed using the $25\times 25\times 1$ $k$-grid.

\subsubsection*{Fe-DCA/Au(111).} It was simulated using the experimentally found $(4\sqrt{3}\times4\sqrt{3})R30^\circ$ structure \cite{Lobo_2023}, with the Fe atoms residing in the substrate's fcc hollow sites. The optimized Au bulk lattice parameter $a_0=2.915$ \AA\, was used to construct the in-plane $(4\sqrt{3}\times4\sqrt{3})R30^\circ$ cell (note that this is the periodicity with respect to the substrate, while for the Fe-DCA itself it is $(1 \times 1)$ honeycomb). This corresponds to the Fe-Fe distance of 11.66 \AA. Three Au layers were used to simulate the Au(111) substrate, so that the Fe-DCA/Au(111) cell contained 224 atoms. The atoms of the lowermost Au layer were fixed during the structural relaxations, while the other two as well as the Fe-DCA layer were allowed to relax. The complex non-co-planar geometry that Fe-DCA acquires when placed on top of Au(111) is described in detail in Ref. \cite{Lobo_2023}. The static total-energy calculations were performed using the $7\times 7\times 1$ $k$-mesh.

\subsubsection*{MnBi$_2$Te$_4$.} The in-plane lattice constant
of the MnBi$_2$Te$_4$ septuple-layer-thick film\cite{Otrokov_2019} was fixed to that of the bulk material ($a_0 = 4.336$ \AA\ \cite{Otrokov_2019, Otrokov_Nat_2019}) to construct the rectangular $(1 \times \sqrt{3})$ cell. The interlayer distances were optimized. The static total-energy calculations were done using a $k$-point grid of $33\times 21\times 1$.

\subsection*{Extracting the model parameters $J,\;D$, and $\lambda$.}
Based on an energetic analysis of the solutions, one can obtain the magnetic anisotropy energy, as well as the isotropic and anisotropic exchange coupling between neighbour spins localized at the 3d magnetic atoms (Co, Cr, Fe and Mn) with different spins ($S=1, 3/2, 2$, and 5/2, respectively). For the four systems under study, we find that they present out-of-plane anisotropy and favorable ferromagnetic coupling between the 3$d$ magnetic atoms' spins. In first approximation, we can obtain the nearest neighbor (isotropic and anisotropic) exchange coupling strengths and single-ion anisotropy from total energy differences of magnetic configurations with parallel and anti-parallel spins along directions perpendicular and parallel to the plane that contains the magnetic atoms~\cite{Lado_Rossier_2dmat_2017}.

\section*{Spin wave theory}
Let us revise the spin wave theory applied to the following spin Hamiltonian describing a ferromagnet in the presence of a homogeneous applied magnetic field. We will derive the energy of different non-interacting spin waves. We first notice that the doubly-degenerate ground state of the system corresponds to a state with all spins aligned.

\subsection*{Single magnon spin waves\label{SW1}}
If we restrict ourselves to low-energy excitations, the deviations of the different spins from the $z$-axis will be small, and one can treat them using spin-wave theory~\cite{Yosida}. A spin wave with minimal spin deviation with respect to the Weiss state can be written as a linear combination of states where the $n$-state is flipped, Eqs. (\ref{SW-l}) and (\ref{SW-q}) in main text.
Thus, we are looking for states of the form 
$|{\rm SW}_1\rangle=\sum_n A_n S_n^+|G\rangle$,
with the normalization condition $\sum_n 2S|A_n|^2=1$. We assume that the exchange integrals $J_{ij}^z$ and  $J_{ij}^\perp$  depend only on the difference between the spins positions ${\bf R}_j-{\bf R}_i$, so that for the spins located on an infinite lattice (in any dimensions), the Hamiltonian has translational symmetry. Thus, each eigenstate of this Hamiltonian must be a basis for an irreducible representation of the translational group of the lattice~\cite{Majlis_book_2007}, so that it satisfies Bloch theorem. Hence, the $|{\bf q}\rangle$ state (\ref{SW-q}) is an eigenstate of $H$. 
It is worth mentioning that for finite periodic systems, i.e., a ring in one dimension or a torus in two dimensions, the situation is fully analogous to the infinite lattice limit with the exception of the quantization of the momentum.\footnote{For a one-dimensional ring with inter spin distance $a$, with the $N+1$-site equal to the $1$-site, the momentum $q$ can take the values $q_n=2\pi n/aN$, with $N=0,\dots, N-1$.}
Taking into account the following relations for $i\ne j$
\beqaSM
S_i^- S_j^+ S_n^+|G\rangle &=& 2S\delta_{i,n}S_j^+|G\rangle,
\crcr
S_i^+ S_j^- S_n^+|G\rangle &=& 2S\delta_{j,n}S_i^+|G\rangle,
\crcr
S_i^z S_j^z S_n^+|G\rangle &=& S^2S_n^+|G\rangle -S \delta_{i,n}S_n^+|G\rangle -S\delta_{j,n}S^+_n|G\rangle,
\crcr
(S_i^z)^2 S_n^+|G\rangle &=& S^2S_n^+|G\rangle +\left(1-2S\right) \delta_{i,n}S_n^+|G\rangle,
\eeqaSM
one can evaluate the action of all the terms in our Hamiltonian on the $|{\bf q}\rangle$ state. In fact, one can demonstrate that\cite{Majlis_book_2007,Nolting_book_2009}
\beqSM
H|{\bf q}\rangle=\left(E_0+E_{SW}({\bf q})\right)|{\bf q}\rangle
\eeqSM
where the energy of the Weiss state can be written as
\beqSM
E_0=-N\left[S^2J^\perp(0)+DS^2+g\mu_B BS\right],
\eeqSM
and the excitation energy $E_{SW}({\bf q})$ with respect to the ground state is given by Eq. (\ref{esmagnon}) in the main text.

\subsubsection*{Honeycomb lattice. \label{honey2D}}
In a honeycomb lattice there are only three first neighbors for each lattice site, and the single-magnon spin wave has two possible solutions to $E_0+E_{SW}({\bf q})$~\cite{Lado_Rossier_2dmat_2017}
\beqSM
E_0^{\rm honeyc}=-N\left[3S^2 J^\perp+DS^2+g\mu_B BS\right],
\eeqSM
and
\beqSM
E_{SW}^{\rm honeyc}({\bf q})=\varepsilon_0 
\pm J^ \perp S\left| f({\bf q})\right|,
\label{esmagnonHc}
\eeqSM
where $\varepsilon_0=3SJ^\perp /\xi+D\left(2S-1\right)+g\mu_B B$ and  $f({\bf q})=1+e^{i{\bf q}\cdot {\bf a_1}}+e^{i{\bf q}\cdot {\bf a_2}}$ is the form factor of the honeycomb lattice, with ${\bf a_1}$ and ${\bf a_2}$ the lattice vectors of the triangular lattice.

\subsubsection*{2-dimensional triangular lattice. \label{hexagonal}}
In the case of the 2d-triangular lattice, there are six first neighbors lying on the corners of a regular hexagon around each of the lattice sites, and only one element in the crystallographic basis. 
Introducing the lattice vectors ${\bf a}_1=a \hat i$ and ${\bf a}_2=a/2\hat i+a\sqrt{3}/2\hat j$, the single magnon excitation energy is then given by
\beqSM
E_{SW}^{\rm trian}({\bf q})=\varepsilon_0^{t}+J^\perp S\left[6/\xi-f^{\rm t}({\bf q})\right],
\label{SW-triag}
\eeqSM
with $\varepsilon_0^{t}=D\left(2S-1\right)+g\mu_B B$ and 
\beqaSM
f^t({\bf q})&=&2\left[\cos({\bf q}\cdot {\bf a}_1)+
\cos({\bf q}\cdot {\bf a}_1)+
\cos\left({\bf q}\cdot ({\bf a}_1-{\bf a}_2)\right)\right].
\crcr
&&
\eeqaSM

\subsubsection*{Periodic chain: ring. \label{ring}}
A periodic one-dimensional chain is in essence a closed ring. The only difference with respect to the infinite chain is the momentum quantization as a result of the boundary conditions. This leads to $q_n= 2\pi n/(aN),\; n=0,\dots, N-1$, where $a$ is the distance between spins in the chain. The single-magnon energy $\epsilon_q$ in Eq. (\ref{esmagnon}) for a ring with first-neighbor exchange $J^{z,\perp}$
 then reads as
\beqaSM
E_{SW}^{\rm ring}(q)&=&2SJ^\perp \left[1/\xi-\cos(qa)\right]+D\left(2S-1\right)+g\mu_B B.
\crcr&&
\label{esmagnonR}
\eeqaSM

\subsection*{Two-magnon spin wave\label{twomagn}}
Here we just recapitulate the main results of Tonegawa~\cite{Tonegawa_ptp_1970}(adapting the notation). 
Let us introduce the following dimensionless parameters:
\beqaSM
\delta &=&\frac{2D(2S-1)\xi}{4J^\perp S}
\\
\nonumber
\delta'&=&\frac{2D\xi }{4J^\perp S}
\\
\nonumber
1/\alpha &=& 4S.
\label{parameters}
\eeqaSM

First, the two non-interacting magnons give place to an energy continuum 
\beqaSM
E_{\rm 2SW}\left({\bf Q},{\bf q}\right)&=&
2SJ^\perp\Big[2/\xi-\sum_j\cos\left(\frac{{\bf Q}\cdot{\bf r}_j}{2}\right)\cos\left({\bf q}\cdot {\bf r}_j\right)\Big]
\crcr
&+&2D(2S-1)+2g\mu_B B,
\label{TMcont}
\eeqaSM
 where the sum is over the first neighbors of the lattice and ${\bf Q}$ is the total momentum of the pair of magnons and whose wavevectors  are ${\bf k_1}={\bf Q}/2+{\bf q}$ and ${\bf k_2}={\bf Q}/2-{\bf q}$. Thus, for a one-dimensional ring one finds an expression that equals twice the energy of a single magnon, i.e.,  
 \beqaSM
E_{\rm 2SW}^{\rm ring}\left(Q,q\right)&=&
4SJ^\perp\Big[1/\xi-\cos\left(\frac{Q a}{2}\right)\cos\left(qa\right)\Big]
\crcr
&&\quad+2D(2S-1)+2g\mu_B B.
\label{TMcont-ring}
\eeqaSM

 In addition, there are two solutions below this energy continuum that correspond to bound states with localized wavefunctions of the two spin excitations whose wavefunctions can be written as
 \beqSM
 |\psi\rangle=\frac{1}{\sqrt{N}}\sum_{ll'}a(l,l')S^+_lS^+_{l'}|G\rangle,
 \eeqSM
 with $N$ the total number of spins. Thus, the amplitude $\psi(l,l')$ for finding the two-spin deviations at the $l$ and $l'$ sites will be given by $4Sa(l,l')/\sqrt{N}$ when $l\ne l'$ and  $2\sqrt{S(2S-1)}a(l,l')/\sqrt{N}$ when $l= l'$~\cite{Tonegawa_ptp_1970}. 
 When $\cos\left({\bf Q}\cdot{\bf r}_j\right)=0$, one of these solutions corresponds to the single-ion type two-magnon bound state, with $\psi(l,l)\ne0$ and $\psi(l,l')=0$ for $l\ne l'$. This solution is known as ``Ising type'' bound state (this bound state only appears for $S>1/2$). The other solution corresponds to a two magnon wavefunction where $\psi(l,l')\ne 0$ for $l'$ a first neighbor of $l$, and $\psi(l,l')=0$ it is known as ``Bethe type'' bound state. For a general ${\bf Q}$, this simple separation is no longer possible.

The energy of the two-magnons bound state $E_{\rm 2M}({\bf Q})$ for a one-dimensional chain 
$E_{\rm 2M}(Q)=4J^\perp S\epsilon/\xi$ 
can be written in terms of the solution to the following cubic equation in $\epsilon$
\beqaSM
(1+\delta-\epsilon)^3+p_2(1+\delta-\epsilon)^2+p_1(1+\delta-\epsilon)+p_0=0
\nonumber
\eeqaSM
where
\beqaSM
p_0&=&-\left[ \left((1-2\alpha)\xi^2\cos^2(Qa/2)-2\alpha\delta'\right)\right.
\crcr
&&\qquad \left.+\delta'^2\xi^2\cos^2(Qa/2)\right]/4\alpha,
\crcr
p_1&=&-(1-2\alpha-\delta')\xi^2\cos^2(Qa/2)+\delta'(2\alpha+\delta')
\crcr
p_2&=&\left[(1-4\alpha)\xi^2\cos^2(Qa/2)-4\alpha(2\delta'+\alpha)\right]/4\alpha,
\nonumber
\eeqaSM
with $a$ the lattice constant.

\end{document}